\documentclass[10pt,letterpaper]{article}

\usepackage[numbers]{natbib}

\usepackage{amsmath} 
\usepackage[version=4]{mhchem}	

\usepackage{graphicx}
\usepackage[export]{adjustbox}
\usepackage[utf8]{inputenc}
\usepackage{nameref,hyperref}
\usepackage[left]{lineno}
\usepackage{microtype}
\DisableLigatures[f]{encoding = *, family = * }

\usepackage{setspace} 
\usepackage{changepage}

\usepackage[aboveskip=1pt,labelfont=bf,labelsep=period,singlelinecheck=off]{caption}

\makeatletter
\renewcommand{\@biblabel}[1]{\quad#1.}
\makeatother

\usepackage{lastpage,fancyhdr,graphicx}
\usepackage{epstopdf}
\usepackage{color}
\definecolor{Gray}{gray}{.25}
\definecolor{red}{cmyk}{0,0,0,1}
\usepackage{graphicx}
\usepackage{sidecap}
\usepackage{wrapfig}
\usepackage[pscoord]{eso-pic}
\usepackage[fulladjust]{marginnote}
\reversemarginpar

\begin{document}
\vspace*{0.35in}
\begin{flushleft}
{\Large
\textbf\newline{Circular spectropolarimetric sensing of vegetation in the field; possibilities for the remote detection of extraterrestrial life}
}
\newline
\\
C.H. Lucas Patty\textsuperscript{1,2,3,*},
Inge Loes ten Kate\textsuperscript{4},
Wybren Jan Buma\textsuperscript{5},
Rob J.M. van Spanning\textsuperscript{1},
G\'abor Steinbach\textsuperscript{6},
Freek Ariese\textsuperscript{7},
Frans Snik\textsuperscript{8}

\bigskip
\tiny 1 Amsterdam Institute for Molecules, Medicine and Systems (AIMMS), VU Amsterdam, De Boelelaan 1108, 1081 HZ Amsterdam, The Netherlands
\\
{2} Institute of Plant Biology, Biological Research Centre, Hungarian Academy of Sciences, P.O. Box 521, H-6701 Szeged, Hungary
\\
{3} Biofotonika R\&D Ltd, Szeged, Hungary
\\
{4} Faculty of Earth Sciences, Utrecht University, Budapestlaan 4, 3584 CD Utrecht, The Netherlands
\\
{5} HIMS, University of Amsterdam, Science Park 904, 1098 XH Amsterdam, The Netherlands
\\
{6} Institute of Biophysics, Biological Research Centre, Hungarian Academy of Sciences, P.O. Box 521, H-6701 Szeged, Hungary
\\
{7} LaserLaB, VU Amsterdam, De Boelelaan 1083, 1081 HV Amsterdam, The Netherlands
\\
{8} Leiden Observatory, Leiden University, P.O. Box 9513, 2300 RA Leiden, The Netherlands

\bigskip

\bf *patty.lucas@brc.mta.hu

\end{flushleft}

\section*{Abstract}
{\color{red}Homochirality is a generic and unique property of all biochemical life and the fractional circular polarization of light it induces therefore constitutes a potentially unambiguous biosignature.} However, while high-quality circular polarimetric spectra can be easily and quickly obtained in the laboratory, accurate measurements in the field are much more challenging due to large changes in illumination and target movement. In this study we have measured various targets in the field, up to distances of a few kilometers, using the dedicated circular spectropolarimeter TreePol. We show how photosynthetic life can readily be distinguished from abiotic matter. {\color{red}We underline the potential of circular polarization signals as a remotely accessible means to characterize and monitor terrestrial vegetation, e.g. for agriculture and forestry. Additionally, we discuss the potential of circular polarization for the remote detection of extraterrestrial life.}

\section{Introduction}

Exoplanetary science has been advancing rapidly over the last few decades. {\color{red}Estimates are that on average with every star of the 100-400 billion stars in our galaxy there is at least one planet \cite{Cassan2012}.} Additionally, estimates on the occurrence of rocky exoplanets in the habitable zone range from 2 \% - 20 \% per stellar system (\cite{Seager2017} and references therein). One of the primary motivations of exoplanetary science is to find signs of life beyond Earth and therefore substantial research is devoted towards the identification, verification and validation of remotely detectable spectral characteristics of exoplanets as a diagnostic of the presence of life \cite{Grenfell2017, Schwieterman2017, Fujii2017, Schwieterman2018, Walker2017, Seager2016, Patty2018a}.

Research on remotely detectable biosignatures has focused on detecting particular atmospheric constituents such as liquid \ce{H2O}, which indicates possible planetary habitability. Additionally, the simultaneous presence of gases in thermodynamic disequilibrium, such as \ce{O2} and \ce{CH4}, has been investigated \cite{Kaltenegger2007, DesMarais2002}. However, detection of these gases is not free of false-positive scenarios \cite{Domagal-Goldman2014, Schwieterman2016, Harman2015, Wordsworth2014}. Other suggested remotely detectable (surface) biosignatures include the well-known red edge effect, resulting from terrestrial vegetation \cite{Seager2005, Kiang2007}, and pigment signatures resulting from other organisms \cite{Schwieterman2015}, with the risk of possible false-positives by mineral reflectance.

{\color{red}Chiral molecules in their simplest form exist in a left-handed (L-) and a right-handed (D-) version that are not superimposable. Unlike abiotic chemistry, different classes of biological molecules tend to select exclusively only one of these configurations (homochirality).} For instance, all living organisms mainly synthesize amino acids in the L-configuration while sugars are predominantly synthesized in the D-configuration. Homochirality also manifests itself within biological macromolecules and biomolecular architectures. The $\alpha$-helix, for example, a common secondary structure of proteins, is exclusively right-hand-coiled. Homochirality is required for processes ranging from proper enzymatic functioning to self replication. The latter is of importance as it implies the prerequisite of homochirality for life \cite{Popa2004, Bonner1995, Jafarpour2015}. It is likely that homochirality is a universal feature of life and therefore may serve as a unique and unambiguous biosignature. 

{\color{red}When interacting with light, chiral molecules cause optical rotation (which is the rotation of the orientation of the linear polarization plane upon interaction with a sample) and exhibit circular dichroism (which is the differential absorption of left- or right-handed circularly polarized incident light).} These phenomena are most evident upon interaction with polarized light, but can also lead to the circular polarizance of unpolarized light, such as emitted from a star \cite{Kemp1987}. Starlight reflecting off an extraterrestrial body with homochiral molecules and molecular systems on its surface will thus carry this information, which, in principle, can be sensed remotely \cite{,Wolstencroft2004, Pospergelis1969, Sparks2009a, Sparks2009, Patty2018a,Patty2017}. 

Polarimetry is in general advantageous for both the detection and the characterization of exoplanets. {\color{red}Polarimetry enables one to enhance the contrast between the starlight and the very dim light reflected off its planets (which often is very strongly linearly polarized) and thus offers the potential to characterize the atmosphere and surface of an exoplanet \cite{Stam2006, Stam2008, Rossi2018}.} Induced linear polarization can potentially be a biosignature. Both biotic and abiotic matter, however, can create large amounts of linear polarization \cite{Shkuratov2006, West1997}. As such, linear polarization might offer complementary information in addition to scalar reflectance signatures, but caution should be exercised when using it for distinguishing biotic from abiotic components. 

{\color{red}As homochirality is exclusive to life, the circular polarization it induces constitutes a strong and potentially unambiguous biosignature. While also abiotic matter can create circular polarization (e.g. through multiple scattering), these signals generally are orders of magnitude smaller than those created by biological homochirality and have a much smoother and broader spectral shape \cite{Sparks2009, Pospergelis1969, Rossi2018}.

Circular polarizance is detected by measuring the induced fractional circular polarization of unpolarized incident light. For the simplest molecules, the polarizance will be strongest where there is higher spectral absorbance and the polarizance will be opposite in handedness to that of the absorbing biological molecule. Excitonic interactions further enhance the magnitude of these signals, but a particularly interesting phenomena arises when observing large and dense supramolecular systems and complexes. These systems can induce anomalously large circular polarization signals with optical activity even outside of the absorbance bands, and has been labeled Polymer and Salt Induced (PSI) circular polarization \cite{Bustamante1980, Keller1986, Garab2009}. While the psi-type circular polarizance can thus extend beyond the absorption bands, the polarization signal will still primarily occur where the total absorption is very high, such that it must be detected within a relatively weak signal, one of the primary challenges for remote observation.}

The amplitude of the signal is also strongly wavelength dependent. Amino acids, for instance, are strongly polarizing in the vacuum-ultraviolet region and thin films (500 nm) of (homochiral) alanine can readily reach a polarizance of 0.6$\%$ at 180 nm \cite{Meierhenrich2010}. Outside the water absorption band ($>$190 nm), the maximum polarizance of the same film is, however, only 0.07$\%$. While abiotic materials can also create circular polarization, often through multiple scattering \cite{Martin2016}, the risk of a false positive scenario is smaller. Various minerals consistently show a much weaker and spectrally different circular polarization signal \cite{Sparks2009, Pospergelis1969}. Also simulations of clouds on Earth or Venus show circular polarizance, but this is small and has a much broader spectral shape \cite{Rossi2018}. Additionally, circular spectropolarimetric sensing of the surface of Mars using ground-based telescopes did not reveal any significant signals, confirming a general lack of false positives \cite{Sparks2005}. Needless to say, chiral molecules will need to have a large enough circular polarizance of the same sign and a large enough abundance at a planetary surface in order to be detectable.

{\color{red}On Earth, algae and phototrophic bacteria have the potential to be detected locally, whereas vegetation has a cover widespread enough to display remotely detectable circularly polarizing features from afar. Photosynthesis is one of the most important hallmarks of life on Earth and is the major life process underlying global primary productivity. Photosynthesis evolved soon after the emergence of life itself \cite{DesMarais2000, Xiong2002, Hohmann-Marriott2011}. This early evolution on Earth and the advantages of being able to utilize the radiation from the host star as an energy source support expectations that photosynthesis will likely evolve on other planets as well. In terms of productivity, surface features, and evolutionary drive, photosynthesis would thus constitute a likely target.}

The circular polarization spectra of terrestrial vegetation relates to the absorbance of its pigments \cite{Garab2009}. The circular polarization features with by far the largest magnitude are found around the chlorophyll \textit{a} Q absorption band ($\sim$680 nm). {\color{red}Typically, a split signal is observed, with a negative (left-handed) band at $\sim$670 nm and a positive (right-handed) band at $\sim$690 nm that are relatively independent. This split circular polarization signal is the result of the superposition of bands of opposite sign originating from different chiral macrodomains \cite{Finzi1989, Garab1991, Garab1988a, Patty2018b}.} It has furthermore been suggested that the local alignments of the chloroplasts might affect the spatial variation in circular polarization and could thus affect the overall signal on the leaf and canopy scale \cite{Patty2018b}. Beside its use as a biosignature for the remote detection of extraterrestrial life, the circular polarizance of vegetation may also be informative for vegetation physiology due to its dependency on the molecular architecture \cite{Patty2017}. 

While high-quality circular polarimetric spectra can be obtained in the laboratory \cite{Patty2017, Patty2018c, Patty2018b}, the next key step is to take dedicated polarimeters into the field. {\color{red}Under these circumstances the instruments will have to cope with very dynamic scenes involving large changes and variability in illumination (i.e. in direction, diffuseness and intensity).} Additionally, due to reflection or scattering a target may be linearly polarized at levels $>10 \%$ and, depending on the modulation approach, mitigation of linear-to-circular polarization crosstalk is crucial. 

In the present study we investigate the circular polarizance of vegetation in the field using TreePol, a dedicated circular spectropolarimeter based on a fast ferro-liquid-crystal (FLC) modulation with dual-beam implementation \cite{Patty2017}. We report the detection of circularly polarized biosignatures of various plants. {\color{red} While previous studies have only measured transmission and reflection of single leaves, we show to the best of our knowledge for the first time the results of measurements on whole plants and even canopies measured at large distances.}

\section{Materials and methods}

\subsection{Polarization}
{\color{red}Polarization in general is described in terms of the four parameters of the Stokes vector $\textbf{S}$. With the electric field vectors $E_{x}$ in the x direction ($0^{\circ}$) and $E_{y}$ in the y direction ($90^{\circ}$), $z=0$, $i=\sqrt{-1}$ and with $^*$ representing the complex conjugate, the Stokes vector is given by:} 

\begin{equation}
\textbf{S}=
\begin{pmatrix}
I\\
Q\\
U\\
V\\
\end{pmatrix}=
\begin{pmatrix}
\left\langle E^{}_{x}E^{*}_{x} + E^{}_{y}E^{*}_{y}\right\rangle\\
\left\langle E^{}_{x}E^{*}_{x} - E^{}_{y}E^{*}_{y}\right\rangle\\
\left\langle E^{}_{x}E^{*}_{y} - E^{}_{y}E^{*}_{x}\right\rangle\\
i\left\langle E^{}_{x}E^{*}_{y} - E^{}_{y}E^{*}_{x}\right\rangle\\
\end{pmatrix}=
\begin{pmatrix}
I_{0^{\circ}}+I_{90^{\circ}}\\
I_{0^{\circ}}-I_{90^{\circ}}\\
I_{45^{\circ}}-I_{-45^{\circ}}\\
I_{RHC}-I_{LHC}\\
\end{pmatrix}
\end{equation}
The Stokes parameters $I$, $Q$, $U$ and $V$ refer to intensities thus relating to remotely measurable quantities. The absolute intensity is given by Stokes $I$. Stokes $Q$ and $U$ denote the differences in intensity after filtering linear polarization at perpendicular directions, where Q denotes the difference between horizontal and vertical polarization and U the difference in linear polarization but with a 45$^{\circ}$ offset. Stokes $V$ denotes the difference between right-handed and left-handed circularly polarized light. $I_{0^{\circ}}, I_{90^{\circ}}, I_{45^{\circ}}$ and $I_{-45^{\circ}}$ are the intensities in the respective planes perpendicular to the propagation axis while $I_{LHC}$ and $I_{RHC}$ are the intensities of left- and right-handed circularly polarized light, respectively. If the absolute intensity $I$ is known, the polarization state can thus be completely described by the normalized quantities $Q/I$, $U/I$ and $V/I$. 

\subsection{Spectropolarimetry}
\begin{figure}[!htb]
	\centering 
	\includegraphics[width=0.95\textwidth]{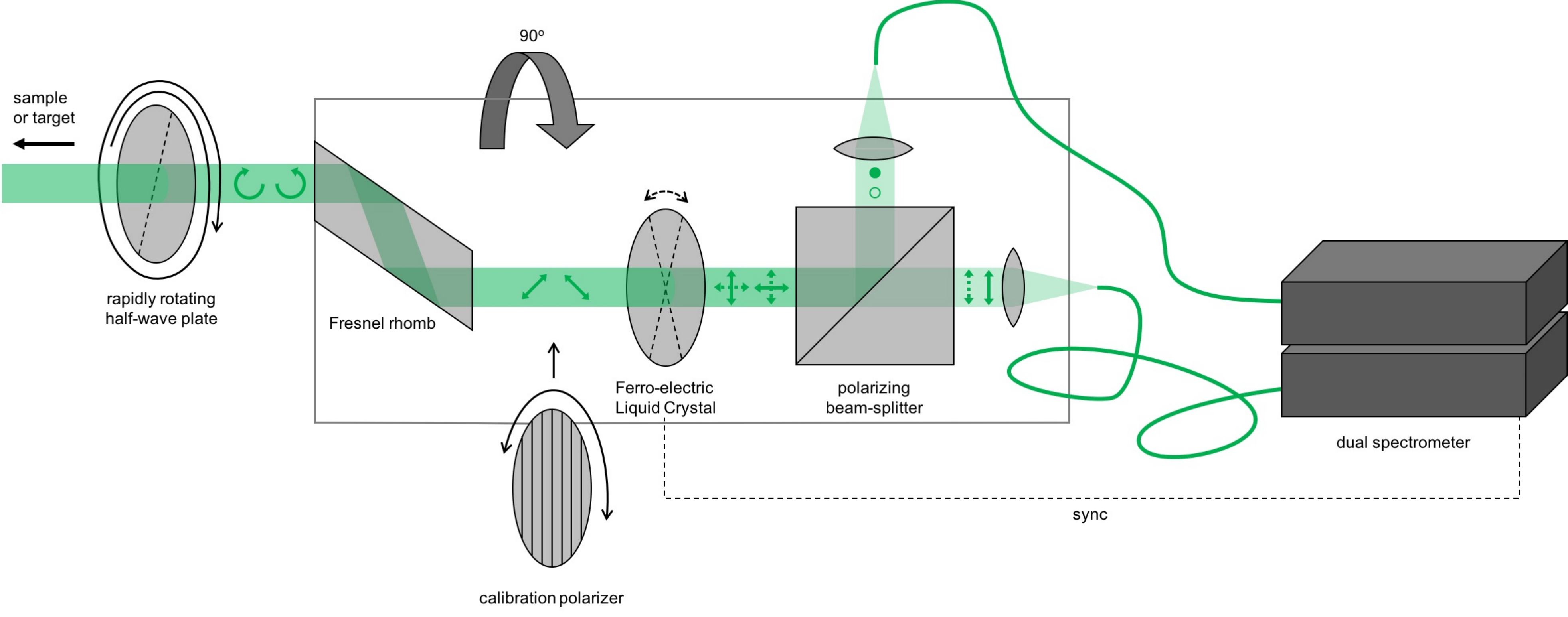}
	\caption{Schematic of TreePol.}
	\label{fig:TreePolCH6}
\end{figure}

{\color{red}Circular polarization measurements, in the lab and in the field, were carried out using TreePol (see also Figure \ref{fig:TreePolCH6} for a schematic) \cite{Patty2017}. TreePol is a dedicated spectropolarimetric instrument developed by the Astronomical Instrumentation Group at the Leiden Observatory (Leiden University), the Netherlands. The instrument was specifically designed to measure the fractionally induced circular polarization ($V/I$) as a function of wavelength (400 nm to 900 nm) and is capable of fast measurements with a sensitivity of $\sim1*10^{-4}$. Treepol measures the fractional induced circular polarization of light after interaction of the sample with unpolarized ambient light. The polarimetric sensitivity is obtained through ferro-liquid-crystal (FLC) modulation, which is synchronized with a dual spectrometer. An achromatic Fresnel rhomb (i.e. a $\lambda/4$ retarder) converts the circular polarization induced by a target into linear polarization which is modulated by the FLC, a $\lambda/2$ retarder with a $45^{\circ}$ switching fast-axis. Additionally, TreePol applies spectral multiplexing with the implementation of a dual-beam approach in which a polarizing beam-splitter feeds the two spectrographs with orthogonally polarized light.

The combination of temporal polarization modulation (i.e. the fast FLC combined with a high-speed spectrometer) with spatial modulation (i.e. simultaneous recording of orthogonal polarization states using two synchronized spectrographs) ensures that systematic differential effects are canceled out to the first order. As such, no spurious polarization signals down to the $10^{-5}$ level are induced \cite{Snik2013}. 
In order to further mitigate possible linear polarization crosstalk, the original design of TreePol (see \cite{Patty2017}) was upgraded with a fast continuously spinning (5 Hz) half-wave plate in front of the Fresnel Rhomb. The angle of view of TreePol is approximately 1.14$^\circ$, which allows for accurate target selection. TreePol was targeted using a calibrated targeting scope mounted on top of the instrument.}

The measurements were carried out with varying integration times, depending on the illumination of the target. Measurements outside were performed using ambient light (i.e. direct sunlight or diffuse light on days with overcast). All measurements were carried out on and around the campus of the Vrije Universiteit Amsterdam, the Netherlands, the Hortus Botanicus Vrije Universiteit Amsterdam, the Netherlands and around the Biological Research Center of the Hungarian Academy of Sciences, Hungary.

\section{Results}
\subsection{Circular Spectropolarimetric measurements}
{\color{red}In previous studies we demonstrated the use of sensitive circular spectropolarimetry in transmission in the laboratory. In the current study we build on the knowledge previously acquired. We will first demonstrate the use on leaves in reflection in the laboratory. We then compare measurements of the same plant measured in the laboratory and outside while trying to mimic the same conditions. Subsequently, we demonstrate the effect (or lack thereof) of ambient light conditions on the circular polarimetric spectra of nearby canopy. Finally, we will show the potential to discriminate between abiotic and biotic matter, even for targets several kilometers away.}

\subsection{Laboratory measurements}
\begin{figure}[htb]
	\centering 
	\includegraphics[width=0.65\textwidth]{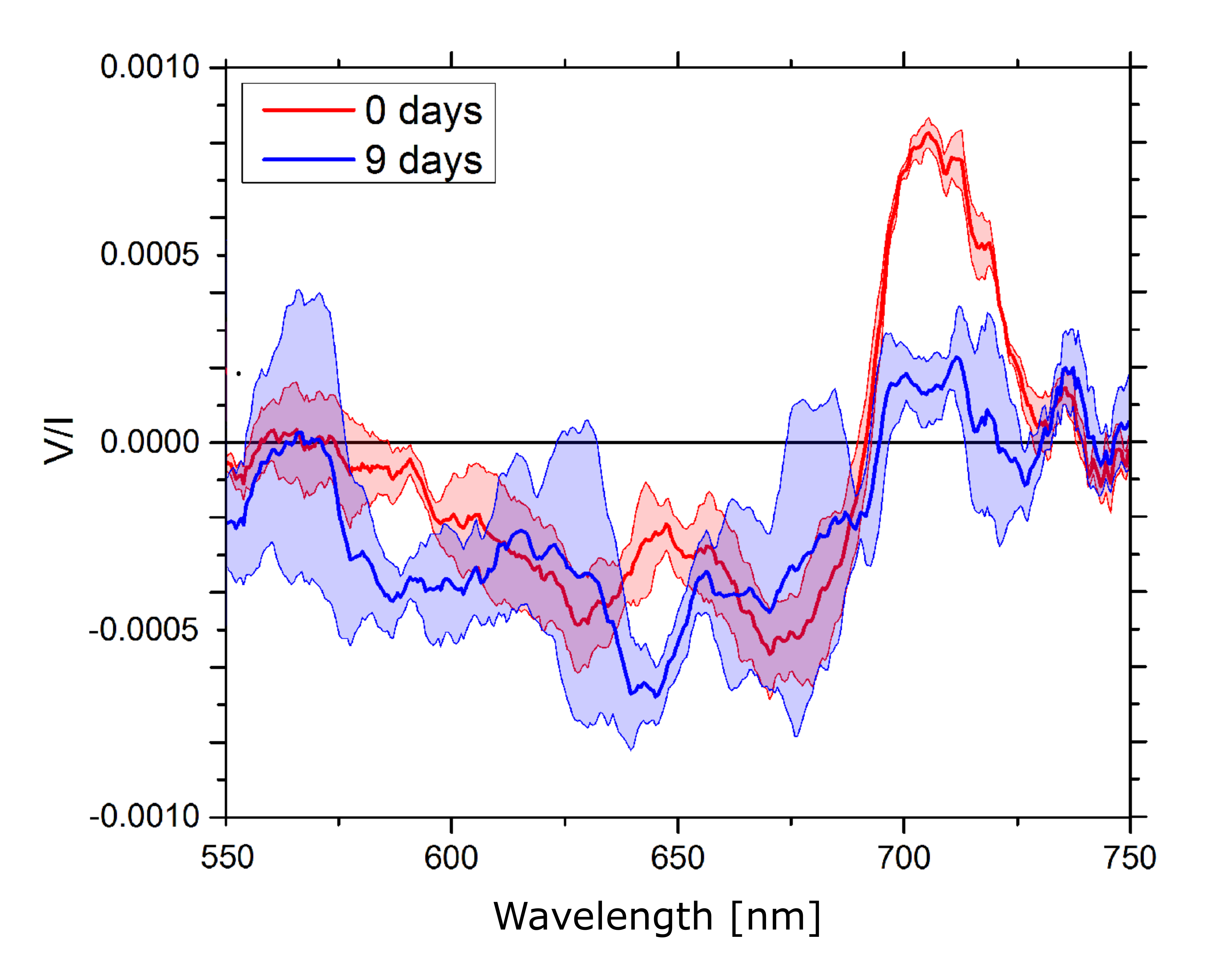}
	\caption{The circular polarimetric spectrum of \textit{Hedera helix} leaves measured in the laboratory using a halogen light source. The same set of leaves (n=3) was measured after 9 days, showing a remarkable reduction of the chiral macrodomains (see also \cite{Patty2017}). Shaded areas denote the standard error.}
	\label{fig:TreePolrefDec}
\end{figure}

In a previous study we used transmission spectropolarimetry to follow the chiroptical signature in leaves that had been cut from their stems and were decaying over time in the dark or under daylight conditions \cite{Patty2017}. These results showed a strong decrease of $V/I$ in time while this decrease was much lower for the chlorophyll \textit{a} concentration. In Figure \ref{fig:TreePolrefDec} we show a small set (same leaves as \cite{Patty2017}) of spectra obtained in \emph{reflection} for leaves stored in the light. These reflection measurements are spectrally very similar to those in transmission, albeit 1 order of magnitude smaller (with the bands for healthy plants showing a maximum amplitude of $+9*10^{-4}$ and $-7*10^{-4}$).

\subsection{Laboratory versus in the field measurements}
\begin{figure}[ht]
	\centering 
	\includegraphics[width=0.95\textwidth]{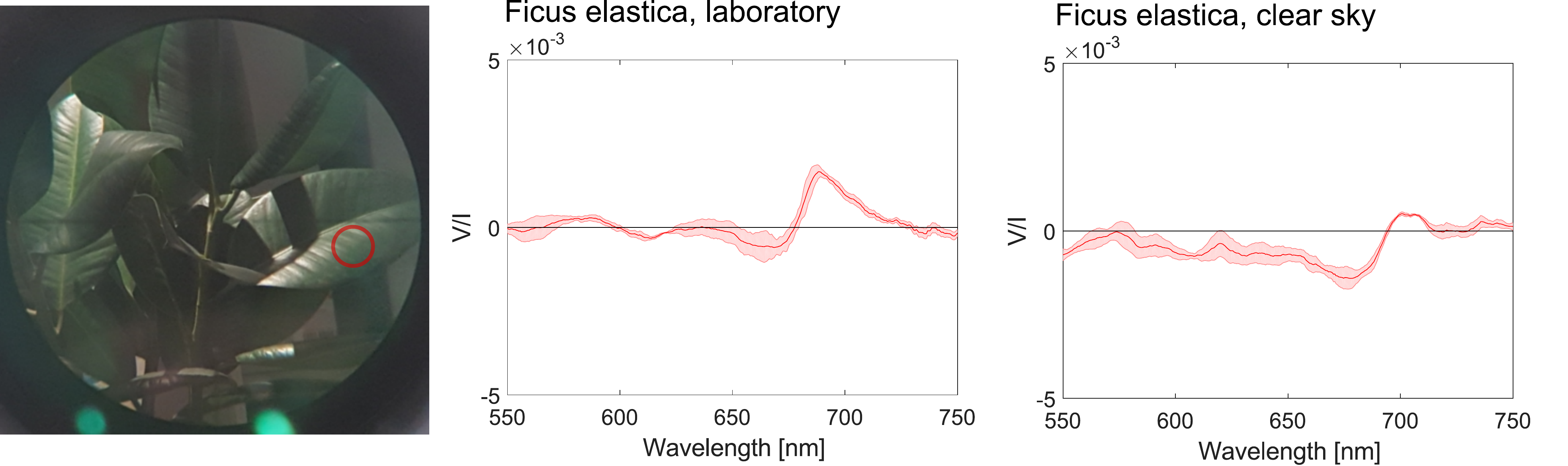}
	\caption{The circular polarimetric spectrum of a \textit{Ficus elastica} measured in the laboratory using a halogen light source (left) and measured under a cloudless overhead sky (right), n=3 for the same area. The red circle (picture not taken at measurement distance) gives the approximate measuring area for the two sets of measurements. The shaded areas denote the standard error.}
	\label{fig:Ficus}
\end{figure}

To further investigate possible differences between laboratory measurements and in-the-field measurements, the circular polarimetric spectrum of an ornamental house plant was measured outside under cloudless conditions and in the laboratory using a halogen light source. The target was measured at a distance of approximately 2 meters (which gives a measurement of a single leaf) while the illumination angle in the laboratory was kept at an approximately equal angle as that of the solar irradiation during the measurements of the same plant outside. {\color{red}Additionally, during the outside measurements any movement due to wind was minimized by a windscreen. Figure \ref{fig:Ficus} shows that inside and outside similar spectral signatures were found, but with some distinct differences: the outside measurements had a wavelength-independent shift to negative values and exhibited a band at 690 nm that was less pronounced.}

\subsection{Ambient light conditions}
\begin{figure}[!htb]
	\centering 
	\includegraphics[width=0.95\textwidth]{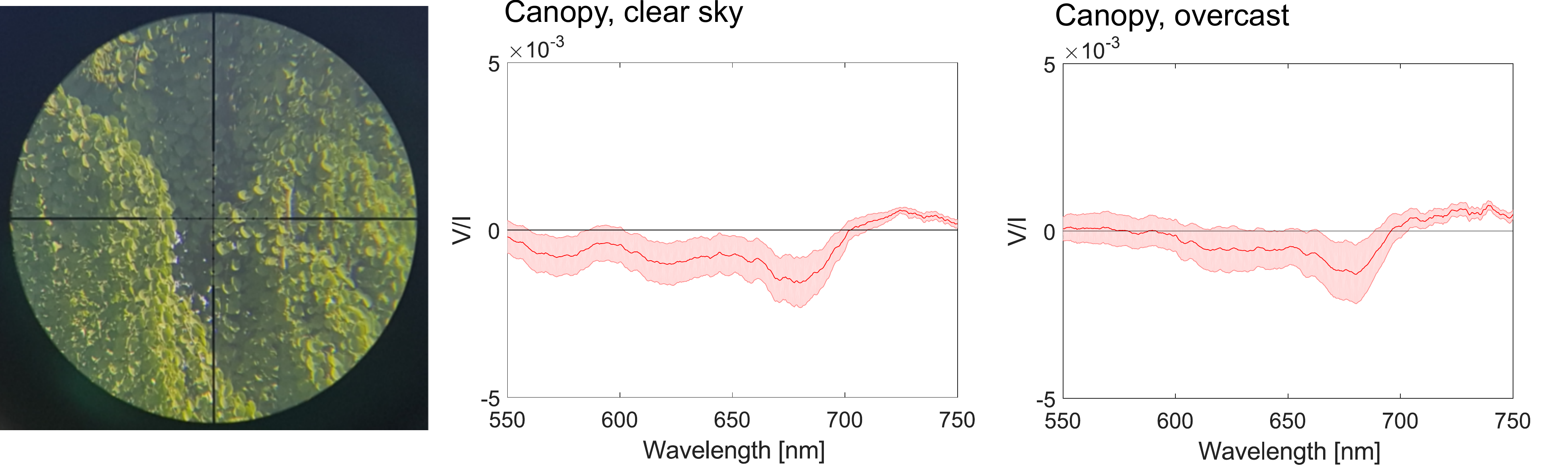}
	\caption{Circular polarimetric spectrum of an unidentified tree measured under a cloudless overhead sky (left) and under overcast conditions (right), n=3 for different parts of the same tree. The measured radius was approximately 0.5 meter. Shaded areas denote the standard error.}
	\label{fig:Suncast}
\end{figure}

The circularly polarizance of vegetation is directly dependent on the photosynthetic machinery located within the cell's chloroplasts, but illumination conditions are expected to have an effect on the magnitude of the signals. We have therefore investigated the effect of light conditions on a single tree. {\color{red}Shown in Figure \ref{fig:Suncast} are the circular spectropolarimetric measurements of the same tree taken under a cloudless sky (left) and under overcast conditions (right).} These measurements were taken at noon and eight days apart from each other. Interestingly, the results are almost identical, indicating that at least for this particular measurement the angle of incidence has no effect. On both days, wind conditions were comparable and below 2 bft. 

\subsection{Biotic versus abiotic matter}
\begin{figure}[!htb]
	\centering 
	\includegraphics[width=0.95\textwidth]{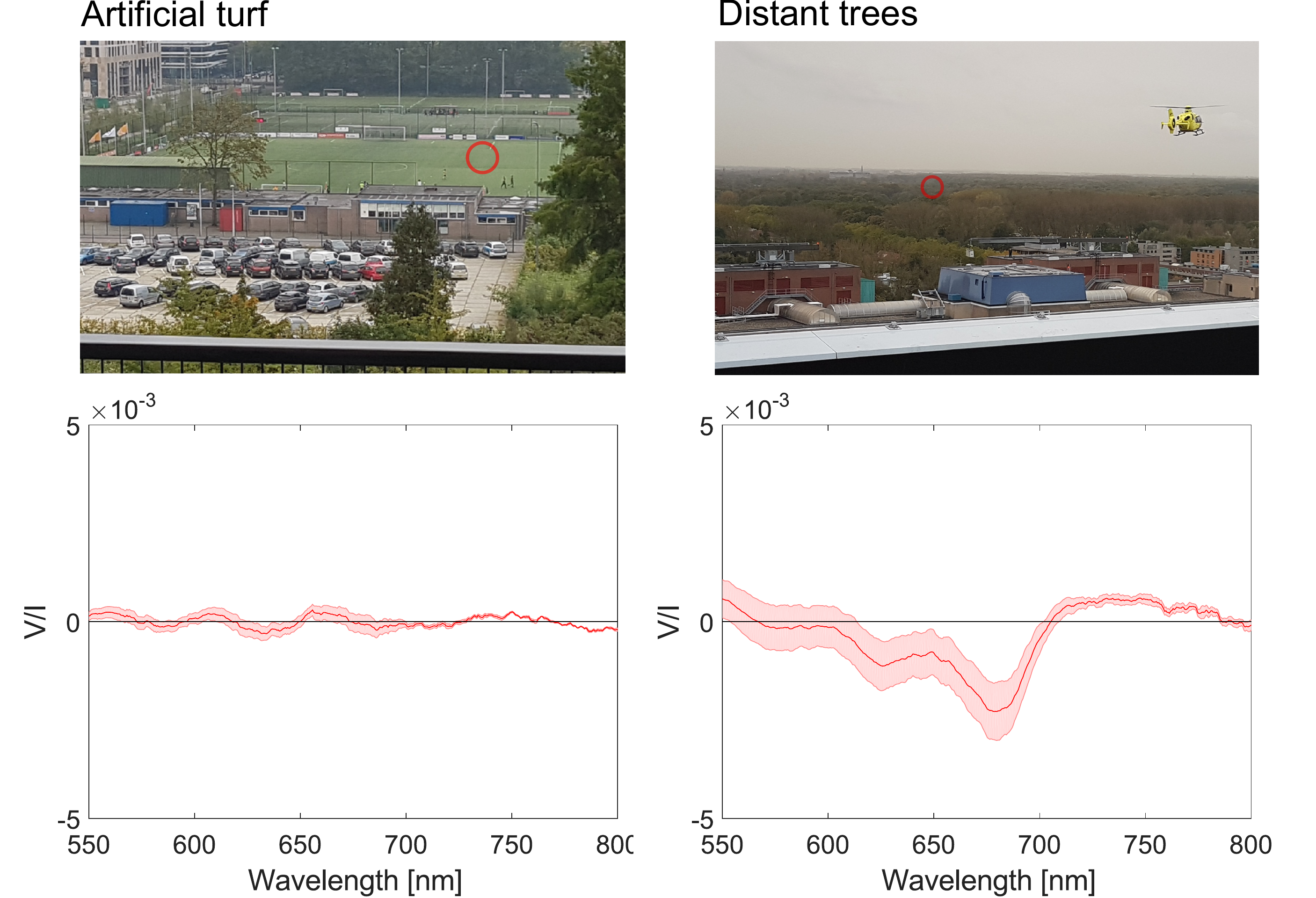}
	\caption{The circular polarimetric spectrum of artificial turf (left) versus that of distant trees (right). The red circle (with a difference between the two photos due to cropping) provides the approximate measuring area. The same area was measured three times, shaded areas denote the standard error.}
	\label{fig:FgrassTrees}
\end{figure}

We have measured several distant targets from the roof of one of the laboratory buildings on the campus of the Vrije Universiteit Amsterdam. Two of these measurements are shown in Figure \ref{fig:FgrassTrees}. The lower left panel shows that the circular spectropolarimetric measurements of a local sports field, consisting of artificial turf/grass, yield no net polarization. The slight deviation ($<5*10^{-4}$) around the zero point shows leftover fringes resulting from the FLC (see also the discussion). Measurements on the tree canopies of the nearby park, 'het Amsterdamse Bos' shown in the right panels, however, show a clear non-zero polarization signal ($\sim 3*10^{-3}$).

\subsection{Field measurements in general}

\begin{figure}[htb]
	\centering 
	\includegraphics[width=0.85\textwidth]{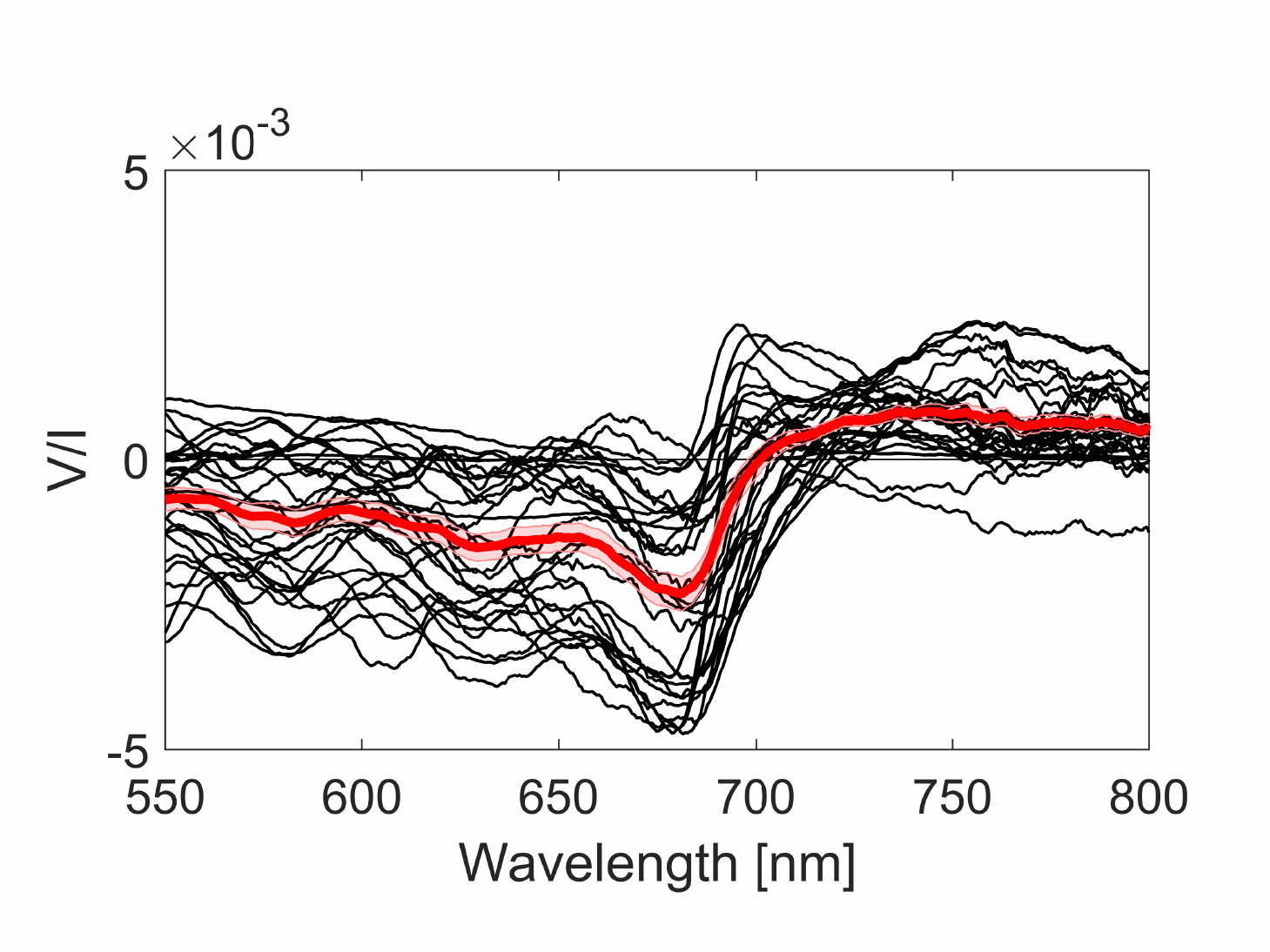}
	\caption{An ensemble of randomly selected outside circular polarimetric spectra of highly diverse vegetation targets, n=30. Measurements were taken of different targets under different ambient conditions. The red graph denotes the mean, the shaded area denotes the standard error.}
	\label{fig:lots}
\end{figure}

An ensemble of various (n=30) measurements of very diverse shrubs and trees in the field and their mean (in red) is shown in Figure \ref{fig:lots}. Included in these results are trees native to the Netherlands and Hungary measured from the rooftops, but also various more exotic shrubs measured at the Hortus Botanicus Vrije Universiteit. Especially the positive band of these reflection measurements is generally smaller than measured in transmission in the laboratory. In transmission, vegetation shows a negative circular polarizance band (at $\sim$ 660 nm) of $\sim -3*10^{-3}$ and a positive band at $\sim$ 680 nm of $\sim 8 * 10^{-3}$.

\clearpage

\section{Discussion and conclusions}

We show here, as far as we know for the first time in the refereed literature, the systematic investigation of circular spectropolarimetric sensing of vegetation in the field. Using our current setup, the circular polarizance of biotic matter can be readily distinguished from that of abiotic matter over a distance of up to several kilometers. This demonstrates the potential of circular polarization spectroscopy for detection of signatures of life. While edges in scalar reflectance spectra and signatures in linear polarization can readily emerge in the spectrometry of abiotic matter, circular spectropolarimetry does not suffer from such drawbacks and is thus able to provide an unambiguous means to detect biotic matter. Consequently, although the circular polarization signals are small, they are a much more exclusive biosignature as they are directly linked to the molecular complexity of life.

The TreePol instrument has been designed especially for highly sensitive and accurate circular spectropolarimetric measurements in the field. One of the limitations to its polarimetric performance is the temporal variation of the scene (i.e., variability in illumination, moving targets, etc.). To overcome these effects, a fast ferroelectric liquid crystal (FLC) based modulation was implemented. Both the switch angle and the birefringence of FLC's are, however, temperature dependent. As such, FLC's can introduce systematic effects like polarized spectral fringing. Usually, these effects can be accounted for by recalibration, but they can not completely be removed (also due to temperature dynamics). They are therefore, for instance, still visible in the measurement of the artificial turf (Figure \ref{fig:FgrassTrees}) and limit the polarimetric sensitivity.

Importantly, although linear polarization cross-talk mitigation strategies are employed, it may not be possible to eliminate these effects completely, especially in a highly dynamic environment. Moreover, while most environmental linear polarization features are relatively spectrally broad, those produced by vegetation (see for example \cite{Peltoniemi2015, Vanderbilt1985, Vanderbilt1985a, Vanderbilt2017, Grant1993}) are essentially inversely related to the red-edge and can thus show steep features around the chlorophyll absorbance band. Obtaining information on the linear polarization properties of the target will be beneficial, even if only for calibration purposes.

The circular polarization signals from light scattering are smaller than those obtained in transmission spectropolarimetry. While the negative band is similar (on average $\sim 3*10^{-3}$), we found that the positive band is on average one order of magnitude smaller ($ \sim 7*10^{-4}$). Importantly, while in transmission most leaves have a very similar magnitude of the circular polarizance normalized by the light intensity, we found much more variation in signals between plants measured under the same conditions in reflection (see also Figure \ref{fig:lots}). A possible explanation may involve the variation in leaf optical properties and the orientation of the leaves.

The spectral characteristics in some of the measurements appear to have a different spectral shape compared to the measurements taken in transmission (cf. \cite{Patty2017}). One of the most noticeable features is the much lower intensity of the largest positive $V/I$ band usually observed. This positive band was more pronounced in the \textit{Ficus} target, both measured in the laboratory and measured outside, than in the deciduous forest canopy. In these measurements, however, the former target had been positioned relatively close to the polarimeter (approximately 1.5 meter), which allowed the measurements to be conducted on single leaves (large leaves; approximately 20$\times$12 cm). As such, rather than observing a canopy with more random and variable orientations, these measurements monitor only one static orientation. At this point we have not yet systematically investigated the influence of solar zenith and azimuth relative to the leaf surface. 

We expected large spectropolarimetric differences depending on the light conditions, because it has been demonstrated that the angle of incidence influences the scalar reflectance properties \cite{Kaasalainen2016}. Even if the circular component would not be influenced directly, the total intensity might be variable which would lead to a different value of V/I. We expected that this phenomenon would be most prominent when comparing measurements on sunny days and on days with overcast. On days with overcast conditions, most of the incident light is diffuse and scattered by the clouds as opposed to the conditions on a sunny day. Our results, however, show that the measurements taken under a cloudless sky are virtually identical (see Figure \ref{fig:Suncast}) to those taken with overcast sky. Possibly, the results might vary with leaf glossiness and the angle of measurements. Additionally, as linear polarization is heavily influenced by light conditions and the angle of incidence, the similar results between the measurements on sunny days and days under overcast conditions attest to the effectiveness of our crosstalk mitigation strategies.

Additionally, we have shown circular spectropolarimetric signals for decaying leaves in reflection spectroscopy (Figure \ref{fig:TreePolrefDec}). Similar measurements were already reported in a transmission setup \cite{Patty2017}. The results presented in the present study demonstrate how also in reflection healthy leaves can be distinguished from unhealthy/dying leaves by the change in $V/I$ spectra. This signal, which is unique to vegetation, rapidly decreases over time and shows a more than 3-fold decrease in magnitude for the positive band after 9 days. We argue that this decrease arises because the macrostructures are not actively maintained due to, for example, drought stress, since the water supply to the leaves has been cut off \cite{Patty2017}. This highlights a possible significance of circular polarization spectroscopy as a remote-sensing tool for vegetation and crop monitoring on Earth.

In the context of astrobiology, the circular polarization of biomolecules is a powerful biosignature \cite{Pospergelis1969, Sparks2009, Sparks2009a, Patty2017, Patty2018a, Wolstencroft2004}. Compared to other surface biosignatures there are no significant signals produced by abiotic matter, and therefore no false positives. The results in this study show a vegetation signal level in terms of circular polarizance on the order of $10^{-4} - 10^{-3}$. While promising in terms of robustness, the magnitude of the signal is quite low and therefore detection of these signals from exoplanets will be challenging. This is especially the case when the signal is further diluted such as can be expected in a planetary disk average, i.e., by surfaces with a higher reflection than those creating the circular polarization. To be able to observe a potentially habitable planet and measure the signals in circular polarization with a high enough signal-to-noise ratio, an extremely large space-based telescope is required. At the moment, only the proposed Large UV/Optical/Infrared Surveyor A (LUVOIR-A) \cite{Luvoir2018} meets this requirement. Additionally, such a telescope will need adaptive optics and advanced coronagraphy to suppress the light of the host star to provide a high enough contrast. In view of the scientific return, such an instrument would, however, certainly be worth its investment. 

One of the other factors that should be taken into account is that while the circular polarization signals are more or less similar for various types of terrestrial vegetation, it is unknown what these levels are for the dominant photosynthesic organisms on other planets. Certain terrestrial brown algae, for instance, display signals varying up to three orders of magnitude in strength and displayed signals up to $2*10^{-2}$ \cite{Patty2018c}. The signals can also result from smaller molecular organisations, leading to excitonic polarization, such as displayed by various phototropic bacteria \cite{Sparks2009}. While we cannot predict the molecular and structural organisation of such biomolecules beyond Earth, the here observed lack of false positives in our approach is encouraging and unique. Any significant signal that is observed is therefore very likely to originate from a highly organized molecular assembly that is prevalent enough to be detectable in the first place, thus suggesting the abundance of something organized in terms of homochiral polymers, and hence probably life. 

We have successfully demonstrated the use of circular spectropolarimetry in the field and our results underline the potential significance of circular polarization both as a remotely accessible means of detecting the presence of extraterrestrial life, and as a valuable remotely applicable tool for vegetation monitoring on Earth. An important next step will be to use these results in (exo)planetary models with realistic components such as different surfaces and clouds (e.g. \cite{Rossi2018, Karalidi2012a, Stam2008}), while future laboratory and field studies (such as the utilization of canopy models to account for leaf angle relative to solar zenith angle) should continue to explore the versatility and potential of this technique.

\section*{Acknowledgments}
We dedicate this work to the memory of our dear friend, colleague and the initiator of this project, Dr. Wilfred R{\"o}ling, who unexpectedly passed away on Friday the 25th of September, 2015. 

This work was supported by the Planetary and Exoplanetary Science Programme (PEPSci), grant 648.001.004, of the Netherlands Organisation for Scientific Research (NWO) and partly by the Economic Development and Innovation Operative Programme (GINOP), grants GINOP-2.1.7-15-2016-00713 and GINOP-2.3.2-15-2016-00001 from the Hungarian Ministry for National Economy. 

\nolinenumbers
\clearpage

\bibliography{./Library/Alles}
\bibliographystyle{ieeetr}
\end{document}